\documentclass[sigconf,nonacm]{acmart}
\renewcommand\footnotetextcopyrightpermission[1]{} 
\AtBeginDocument{%
  }

\setcopyright{none}
\acmConference{}{}{}
\acmBooktitle{}
\acmYear{}
\copyrightyear{}
\acmDOI{}
\acmISBN{}

\acmConference[SoCC '25]{Make sure to enter the correct
  conference title from your rights confirmation email}
\acmISBN{}

\usepackage{multirow}

\begin{document}

\title{FC-ADL: Efficient Microservice Anomaly Detection and Localisation Through Functional Connectivity}

\author{Giles Winchester}
\email{G.Winchester@sussex.ac.uk}
\orcid{1234-5678-9012}
\affiliation{%
  \institution{University of Sussex}
  \city{Brighton}
  \country{UK}
}

\author{George Parisis}
\email{G.Parisis@sussex.ac.uk}
\orcid{1234-5678-9012}
\affiliation{%
  \institution{University of Sussex}
  \city{Brighton}
  \country{UK}
}

\author{Luc Berthouze}
\email{L.Berthouze@sussex.ac.uk}
\orcid{1234-5678-9012}
\affiliation{%
  \institution{University of Sussex}
  \city{Brighton}
  \country{UK}
}

\renewcommand{\shortauthors}{Winchester et al.}

\begin{abstract}
Microservices have transformed software architecture through the creation of modular and independent services. However, they introduce operational complexities in service integration and system management that makes swift and accurate anomaly detection and localisation challenging. Despite the complex, dynamic, and interconnected nature of microservice architectures, prior works that investigate metrics for anomaly detection rarely include explicit information about time-varying interdependencies. And whilst prior works on fault localisation typically do incorporate information about dependencies between microservices, they scale poorly to real world large-scale deployments due to their reliance on computationally expensive causal inference. To address these challenges we propose \textit{FC-ADL}, an end-to-end scalable approach for detecting and localising anomalous changes from microservice metrics based on the neuroscientific concept of \textit{functional connectivity}. We show that by efficiently characterising time-varying changes in dependencies between microservice metrics we can both detect anomalies and provide root cause candidates without incurring the significant overheads of causal and multivariate approaches. We demonstrate that our approach can achieve top detection and localisation performance across a wide degree of different fault scenarios when compared to state-of-the-art approaches. Furthermore, we illustrate the scalability of our approach by applying it to Alibaba's extremely large real-world microservice deployment.
\end{abstract}



\maketitle

\section{Introduction}
\label{sec:introduction}
Microservices (MSs) have fast become a cornerstone of contemporary software architecture. Under this paradigm, an application is built as a suite of small, independent services that communicate with each other over well-defined APIs to provide application functionality. Owing to distinct benefits such as developmental agility and fine-grained scaling, MSs are becoming the leading method for orchestrating online services and have already been integrated into many large service providers such as Netflix, eBay, Uber, and Alibaba.

While MS systems outclass traditional monolithic approaches in many aspects, they can also introduce significant operational challenges \cite{noauthor_microservices_nodate, minutes_challenges_nodate}, particularly in regards to meeting acceptable up-time guarantees. With downtimes associated with high costs \cite{chen_empirical_2019}, accurate anomaly detection and localisation is paramount. However, as the systems grow, it becomes increasingly difficult to track the full set of inter-dependencies between components, particularly, as, by design, these relationships are constantly in flux during runtime operation, through many factors such as changing user behaviour, service orchestration, load balancing, and horizontal scaling.

To address these challenges, a number of promising anomaly detection and localisation approaches, with a specific interest on faults, have been developed that seek to leverage the vast amounts of operational data collected from microservice architectures (MSAs) \cite{luo_characterizing_2021}. These approaches can be broadly categorised based on the primary observability data that they leverage into trace-based and metric-based approaches. However, despite the promising performance of many of these approaches, they face significant limitations in practice. Trace-based methods rely on comprehensive MSA instrumentation and primarily detect anomalies that manifest in direct invocation paths, not considering indirect methods for fault propagation such as resource contention \cite{ma_automap_2020}. Metric-based approaches, whilst less intrusive, often treat metrics from MSs independently, overlooking the contribution of complex inter-MS interactions or rely on precise anomaly detection times which are often impractical to obtain in a real-world system \cite{pham_baro_2024}. Recent work has begun to develop metric-based approaches that leverage the inter-dependent nature of MSAs such as through causal structural or multivariate modelling \cite{yang_hi-rca_2023, ma_automap_2020, chen_causeinfer_2019, lin_microscope_2018, qiu_causality_2020, wang_cloudranger_2018, mariani_localizing_2018, wu_microdiag_2021, ma_ms-rank_2019, ma_servicerank_2022, lin_facgraph_2018, xin_causalrca_2023, pham_baro_2024}, however, many of these approaches lack end-to-end anomaly detection and localisation or are computationally expensive, limiting their application to MSAs larger than smaller-scale research testbeds \cite{seshagiri_sok_2022}.

Motivated by the above challenges, in this work we introduce FC-ADL, \textbf{F}unc\-tional \textbf{C}on\-nectivity-based \textbf{A}nomaly \textbf{D}etection and \textbf{L}\-ocalisation. Inspired by work leveraging the concepts of \textit{functional connectivity} (FC) \cite{messager_inferring_2019} or \textit{functional dependencies} \cite{graf_semantic-driven_2022}, we develop a framework to construct a time-varying graph representation of the statistical relationships between MSs at runtime. These time-varying FC dependencies are inferred directly from usage metrics using light-weight correlation measures without requiring distributed tracing data (\ref{sec:inferring}). 
By efficiently quantifying the structural distance between these graphs, we demonstrate that anomalies can be detected directly from changes in these FC-based dependencies (\ref{sec:fc_structural}, \ref{sec:detection}). By focusing on changes in MS relationships rather than individual metric values, FC-ADL is more robust to spurious detection from time-varying metric data. Furthermore, we demonstrate that FC-ADL can accurately locate the candidate root cause of anomalies without needing causality by identifying the MS component with the largest change in FC dependencies before and after a change-point. We validate FC-ADL by comparing its fault detection and localisation performance against state-of-the-art baselines on a new MS fault testbed by integrating chaos engineering tools \cite{noauthor_chaos-meshchaos-mesh_2024} and a time-varying workload generator \cite{noauthor_locustio_nodate} into the widely used DeathStarBench testbed \cite{gan_open-source_2019} which allows for more realistic fault-based detection and localisation testing. We also demonstrate the scalability of FC-ADL by comparing its computational complexity to other detection and root cause analysis (RCA) approaches when applied to a large-scale MS system maintained by Alibaba \cite{luo_characterizing_2021}. An open-source implementation of FC-ADL, the developed fault testbed, and all code and data associated with evaluation will be released alongside this paper \cite{FC-ADL2025}.


\section{Related Work}
\subsection{Trace-Based Anomaly Detection}
Trace-based detection techniques leverage distributed tracing data to infer inter-MS dependencies and detect anomalies, often through structure-similarity methods or machine learning models trained to represent the normal operational patterns of a system's invocation paths \cite{wang_workflow-aware_2020, meng_detecting_2021, huang_tprof_2021, liu_unsupervised_2020, nedelkoski_anomaly_2019, panahandeh_serviceanomaly_2024, meng_midiag_2020, chen_tracegra_2023, shahini_autoencoder-based_2024, zhang_efficient_2023}. These methods can identify anomalous propagation patterns through call graphs and can then be paired with graph-based ranking techniques \cite{yu_tracerank_2023} to localise root causes. However, distributed tracing data is generally limited to anomalies that manifest through changes in invocation chains or invocation slowdown, which, whilst providing valuable insights into how services interact and which dependencies are involved in an anomaly, is limited to the service or endpoint level. Additionally, anomalies within MSAs can manifest due to indirect interactions that are not limited to direct invocation-based propagation. For example an anomaly that propagates through co-located MSs on the same node due to resource contention may be misattributed by tracing-based approaches \cite{ma_automap_2020}. Furthermore, instrumenting and maintaining distributed tracing at scale is difficult and prone to incomplete data collection \cite{sridharan_observability_2018, shen_networkcentric_2023} if possible at all due to ownership and privacy concerns \cite{zhai_nrcac_2025}. What's more, tracing data can be associated with runtime overheads \cite{eder_comparison_2023} and costs in processing and storing large volumes of data \cite{winchester_temporal_2023, luo_characterizing_2021}.

\subsection{Metric-Based Anomaly Detection}
Metric-based approaches rely on component or microservice-level measurements such as resource usage rates, network traffic, and response latency. Unlike trace-based approaches, they can capture anomalies that occur due to indirect dependencies and often do not require the same extensive instrumentation. One set of statistical approaches for anomaly detection from metric data focuses on time-series outlier detection, based on the assumption that anomalies present as unusual metric values \cite{samir_dla_2019, samir_anomaly_2019, vallis_novel_nodate, li_causal_2022, lin_microscope_2018, wu_microdiag_2021, xin_causalrca_2023, lee_eadro_2023, meng_localizing_2020}. However, such approaches equally apply the same assumptions to all metrics, when in reality metrics can show very different behaviours due to the heterogeneity of MSAs. Another approach, involving either statistical \cite{yu_microrank_2021, chen_framework_2020, li_universal_2021} or ML techniques \cite{gulenko_detecting_2018, wu_microdiag_2021, hacid_performance_2021, xu_lightweight_2017, jin_anomaly_2020}, is to model the normal behaviour of individual metrics during anomaly-free runs of the system and use deviations in individual metrics from this normal behaviour to infer anomalies. A key shortcoming of this approach is that as MSAs are highly inter-connected, focusing on metrics from MSs individually can lead to false inference. For example, two microservices might only show sub-threshold changes in their usage metrics despite having stopped interacting due to a fault where the computational overhead of the functionality is low. Conversely, metrics might show spikes in activity due to increases in front-end demand or changing user behaviour without any change in the strength of dependencies between microservices. 

\subsection{Metric-Based Root Cause Analysis}
Once an anomaly has been detected, the next step is localise what caused the anomaly through root-cause analysis (RCA). One set of more simple approaches rely on comparing the statistics of metrics before and after detection, ranking the most likely candidate root causes based on the largest deviations \cite{li_causal_2022, lin_microscope_2018, wu_microdiag_2021, xin_causalrca_2023, lee_eadro_2023, meng_localizing_2020}. However, such approaches do not consider the propagation of faults between MSs and confounding factors such as resource contention on a co-habited node. A popular set of approaches for RCA from MS metric data that consider these factors are causal graph-based analysis approaches \cite{yang_hi-rca_2023, ma_automap_2020, chen_causeinfer_2019, lin_microscope_2018, qiu_causality_2020, wang_cloudranger_2018, mariani_localizing_2018, wu_microdiag_2021, ma_ms-rank_2019, ma_servicerank_2022, lin_facgraph_2018, xin_causalrca_2023}. The common principle behind these approaches is the construction of a causal graph where vertices represent individual MSs or MS metrics, and edges represent the causal relationship between these MSs or metrics. These graphs are commonly constructed using causal structural modelling methods such as PC and LiNGAM \cite{ma_automap_2020, chen_causeinfer_2019, qiu_causality_2020, wang_cloudranger_2018, mariani_localizing_2018, ma_ms-rank_2019, ma_servicerank_2022, lin_facgraph_2018, wu_microdiag_2021, xin_causalrca_2023}. Motivated by the observation that anomalies within MSAs propagate from the original root cause to dependent metrics several different approaches are then taken to locate candidate root causes from these causal graphs, such as applying centrality graph measures \cite{mariani_localizing_2018, wu_microdiag_2021, xin_causalrca_2023}, random walks \cite{ma_automap_2020, wang_cloudranger_2018, ma_ms-rank_2019, ma_servicerank_2022}, or depth or breadth-first searches \cite{chen_causeinfer_2019, lin_microscope_2018, qiu_causality_2020}. However, the time complexity of such causal modelling approaches can be prohibitive. For example, the PC algorithm is worst-case exponential on the number variables as each variable must be tested for conditional dependencies on all subsets of other variables \cite{spirtes_causation_1993, colombo_order-independent_2014}. Variants of LiNGAM do not incur the same combinatorial overhead as the PC algorithm, they still often operate in polynomial time on the number of variables due to matrix operations and independent component analysis steps \cite{shimizu_linear_2006, shimizu_directlingam_2011}. Furthermore, many of these localisation approaches assume the accurate detection of anomaly onset, which can not be guaranteed in complex operating environments \cite{pham_baro_2024}. Moreover, many of these approaches only concern themselves with RCA and do not implement end-to-end anomaly detection and localisation. Recently, BARO \cite{pham_baro_2024} introduced an end-to-end approach that models time-varying changes in multivariate metric time series for end-to-end detection and localisation. However, to update run-length distributions, online determinant and inverse operations, that do not scale well to the number of variables, must be repeatedly carried out.

\begin{figure*}[ht]
    \centering
    \includegraphics[width=0.85\textwidth]{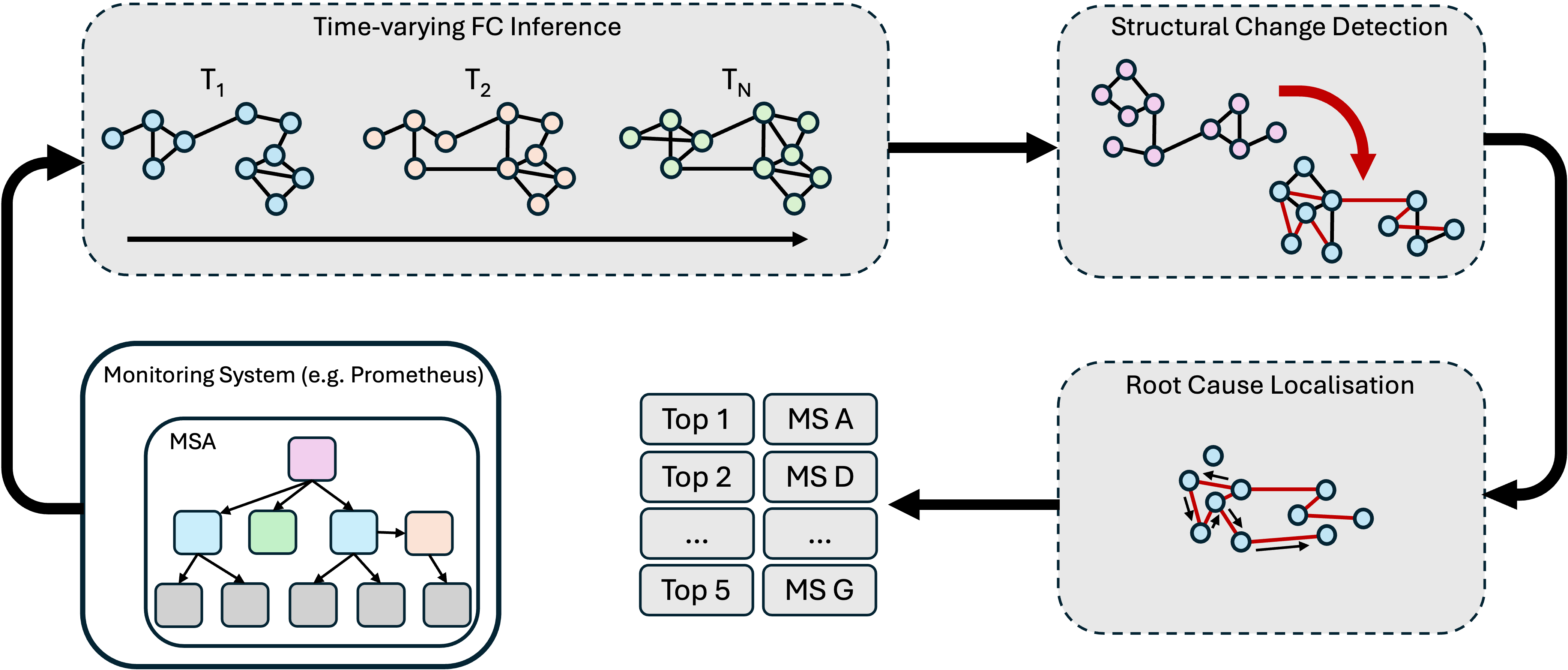} 
    \caption{An overview of the FC-ADL framework}
    \label{fig:framework}
\end{figure*}

\section{Framework}

The design of FC-ADL is motivated by shortcomings of prior work identified above, namely:
\begin{itemize}
\item The use of tracing data which limits the scope of anomalies that can be identified and incurs significant costs for the operator; 
\item The insufficient consideration of the time-varying properties of MS interdependencies that can lead to false inference on the presence of an anomaly;
\item The lack of end-to-end frameworks for anomaly detection and localisation;
\item The reliance of either detection or root-cause analysis on methods that typically do not scale to large real-world deployments; 
\end{itemize}

Depicted in Figure~\ref{fig:detecting_faults}, FC-ADL involves three main processes, time-varying FC inference and structural distance quantification from MS metrics, structural change-based anomaly detection, and FC-based localisation, which we detail below. 

\subsection{Inferring Time-Varying MS Functional Connectivity}
\label{sec:inferring}
In neuroscience, the concept of \textit{functional connectivity} denotes the graph of statistical dependencies among neurophysiological activities, offering insights into functional relationships without requiring knowledge of structural (physical) connectivity. Besides neuroscience, this concept has been shown to be applicable to the analysis of a number of large scale distributed systems, including computer networks \cite{messager_inferring_2019} and systems of systems \cite{graf_semantic-driven_2022}, with further work showing how it can accelerate root-cause analysis \cite{winchester_accelerating_2022}. In what follows, we describe how to extract time-varying FC from correlations between the fluctuations in the metric observability data collected from MSs.

\subsubsection{FC-Based Dependency Graph Construction}
The FC of a system is represented as a graph, $G=(V,E)$, where the set of vertices, $V$, and undirected edges, $E$, represent the components of the system and the strength of statistical dependency between their activities respectively. Within this work we leverage MS-level CPU usage metrics collected using Prometheus for two primary reasons. Firstly, CPU usage is a standard metric exposed by pods, this means that this metric can also be collected in the black box scenario where a cloud provider is monitoring customer workloads, whereas metrics like request latency cannot due to privacy concerns and lack of ownership over exposed metrics \cite{zhai_nrcac_2025}. Secondly, we argue that CPU usage better represents instantaneous MS activity as it directly reflects the processing of work in real time, whereas other standard metrics such as memory usage can remain allocated for extended periods regardless of whether a MS is actively handling requests. As we demonstrate in our results in Sections \ref{sec:anomaly_detection_eval} and \ref{sec:rca_eval}, CPU usage is generally sufficient to detect and localise various non-CPU related faults.

Constructing the FC from the CPU usage of MS systems is non-trivial, as FC-ADL must be able to update on the timescale of seconds even for very large scale MS deployments, necessitating scalability. Moreover, MS CPU usage data is non-stationary, demonstrating strong auto-correlation, seasonality, and trends. This non-stationarity can be handled by sophisticated time series models \cite{bauwens_multivariate_2006, shumway_arima_2017}, however, these approaches often have scalability issues. Therefore, in this work, we rely on Pearson's correlation, which is often used to infer FC in neuroscience \cite{cohen_measuring_2011}, and CPU usage first-order differencing. This combined approach mitigates the effects of non-stationarity whilst leveraging the computational efficiency of calculating Pearson's correlations. To capture time-varying changes in MS dependencies, we generate consecutive temporal snapshots of FC by calculating Pearson's correlations over sliding windows. When each new CPU usage sample is collected from the system, the windows are shifted by one data point to generate a new FC graph. To mitigate the delay in detecting changes in dependencies, we use the exponentially weighted moving average (EWMA) Pearson correlation coefficient, which applies exponential smoothing to correlations calculated within the windows to give more weight to recent observations \cite{pozzi_exponential_2012}.

As measurements from MS metrics are often noisy in nature, a large number of Pearson correlations will be negligible but non-zero. This dense property can mask the true FC structure, and potentially obscure changes in it. Based on the assumption that small correlations are more likely to be associated with noise than true relationships between MSs we apply a nominal threshold in all cases of 0.1. Thresholding or graph filtering are typical steps in constructing stable FC's within neuroscience \cite{friston_functional_2011}.

\subsection{FC Structural Distance}
\label{sec:fc_structural}
To track how the FC between MSs evolves over time we characterise the structural distance between subsequent FC graphs. Here, we follow the methodology described in \cite{masuda_detecting_2019} for detecting sequences of system states in temporal networks but implement DeltaCon as our distance measure rather than the eigenspectrum. We choose DeltaCon over the eigenspectrum due to its better sensitivity to local changes in network structure and improved robustness to noise, which is more suitable when dealing with MS-level faults in potentially noisy environments. In our approach each inferred sliding-window FC after thresholding is treated as a weighted adjacency matrix $A_n$ of a graph $G_n$. To compute the DeltaCon distance we follow the steps outlined in \cite{koutra_deltacon_2013} by computing the node affinity matrix

\begin{align}
\mathbf{S}_n 
&= \bigl(\mathbf{I} + \epsilon^2\,\mathbf{D}_n \;-\;\epsilon\,\mathbf{A}_n\bigr)^{-1} 
\end{align}

where $\epsilon = 10^{-2}$, $D_n = diag(A_n \mathbf{1})$ is the degree matrix, and $I$ is the identity. We then measure the graph-distance between windows $m$ and $n$ via the root-Euclidean distance on these affinities

\begin{align}
d_{m,n}
&= \sqrt{\sum_{i,j}\Bigl(\sqrt{S_{m}^{ij}} - \sqrt{S_{n}^{ij}}\Bigr)^2}
\end{align}

The resulting distance matrix $D$ with entries $d_{m,n}$ captures the structural dissimilarity of FC graphs over time.

\subsection{FC-Based Anomaly Detection}
\label{sec:detection}
In our FC-ADL framework we propose that changes in dependencies between MS, inferred using metric-based FC, can directly be used to determine when an anomaly has occurred, rather than focusing on deviations of a single metric in isolation. To determine when a change point within the FC structure of MSs occurs we apply HDBSCAN \cite{campello_hdbscan_2013} to cluster the DeltaCon distances between FC graphs into communities. HDBSCAN has two beneficial properties for our application, firstly it is density-based and therefore automatically determines the appropriate number of clusters $k$. This means that, unlike other approaches that have been used for MSA anomaly detection \cite{li_causal_2022, shan_diagnosis_2019}, the approach does not require prior knowledge as to whether an anomaly has occurred. Secondly, once fit, HDBSCAN can provide a probability metric, based on membership strength, as to how well new inferred FC-graphs fit into existing clusters. Thirdly, HDBSCAN provides outlier detection which both improves robustness to noisy behaviour of the MSA, but also provides an approach, in conjunction with the probability metric, to provide online re-clustering in response to changing dependencies within the MSA.

\subsection{FC-Based Localisation}
\label{sec:rca}
The principal concept behind RCA within FC-ADL is that structural changes in FC dependencies can also indicate what caused the change. This assumption is rooted in prior RCA research, where causality analysis is used to infer the directed relationships between metrics during an anomaly to identify the root cause \cite{ma_automap_2020, wu_microrca_2020}. However, we propose instead that this same process can be efficiently carried out on undirected relationships inferred based on correlation. To carry out RCA in FC-ADL we first identify the FC dependencies that have changed between the prior normal state of the system and the current abnormal state. To identify these dependencies we average the collection of FC graphs associated with the normal and abnormal state of the system to produce the centroid graphs $G_n$ and $G_a$ that represents the average normal and abnormal FC dependency structure respectively. Next, to isolate the changes in FC that are only associated with the anomalous state of the system we subtract $G_n$ from $G_a$ to give us the dependency change graph $G_c$. We take the absolute value of $G_c$ and apply the same nominal threshold of $0.1$ to remove noisy low-strength edges. The resultant weighted graph $G_c$ denotes which FC dependencies have changed between the two states of the system, and the strength of their change.

To provide a list of candidate root causes, we apply a random walk to $G_c$ to quantify the contribution of each MS to changes in dependencies between the two states. Based on the propagating nature of anomalies within MSAs, we assume that the MS that has the most impact on the structural changes in FC dependencies is more likely to be the root cause of the anomaly. Unlike previous approaches \cite{ma_automap_2020, ma_ms-rank_2019} our random walker implements only a single type of transition probability as the graph is undirected and only forward transitions are required. Unlike causal approaches \cite{wu_microdiag_2021}, FC-ADL inherently captures the strength of edges between vertices, which is incorporated into the walker without further steps required for inferring weights for individual MS. For our random walker, the probability of walking from MS $v_i$ to $v_j$ is proportional to the strength of the edge $e_{i,j}$. We generate random walks from each node in $G_c$ and then sum the visitation frequency of each vertex from all walks. The output of this process is a list of visitation frequency in descending order, with the top-K MS(s) in the list being provided as the most probable root cause(s).

\section{Implementation and Evaluation}
\subsection{FC-ADL Implementation}
The components of FC-ADL have several free parameters, which, for this evaluation, are set as follows. The window size was set to 360 to capture the correlations between 30 minutes worth of CPU usage data, and the step size was set to 1, producing a FC graph for every newly sampled data point. For clustering, we used \textit{scitkit's} HDBSCAN implementation with the minimum cluster size set to $|D|$/5 and all other parameters set to default. For root cause localisation we apply the random walker to generate 10 walks from each vertex in the graph, with a maximum walk-length of 10. Full implementation details can be found in the open-source implementation of FC-ADL released with this paper \cite{FC-ADL2025}.

\subsection{Experimental Datasets}
\label{sec:experimental_setup}
\subsubsection{DeathStarBench}

\begin{figure*}[ht]
    \centering
    \includegraphics[width=0.85\textwidth]{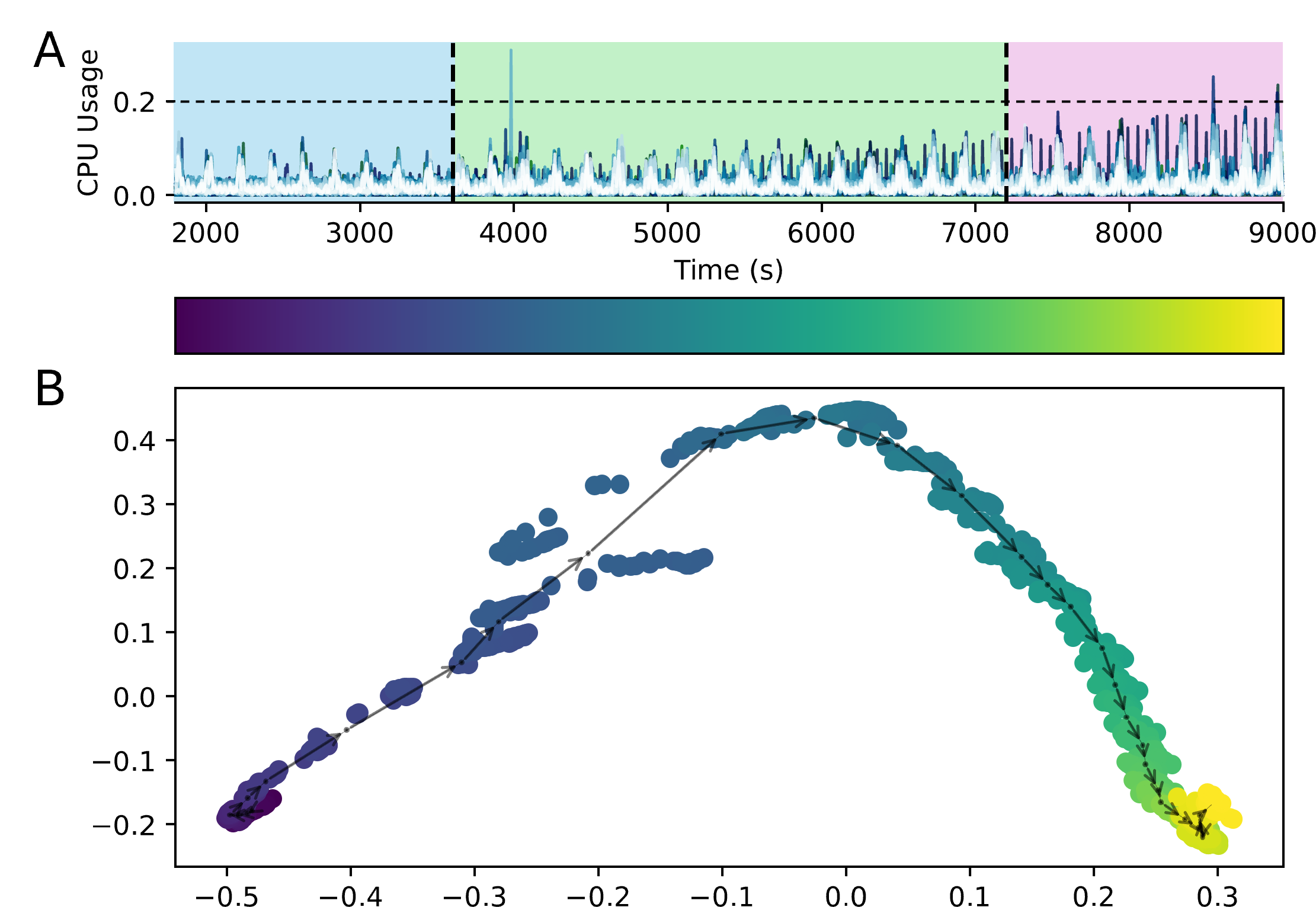} 
    \caption{Panel A: CPU usage for each individual MS instance over the course of the experiment, with the background colour indicating only \textit{read home timeline} user behaviour (blue), time-varying mixed user behaviour (green), and stationary user behaviour with changing amplitude (pink). Panel B: The LMDS 2D embedding of spectral distances between FC graphs, with colour indicating temporal ordering.}
    \label{fig:time-varying}
\end{figure*}

is an open-source benchmark suite for MS deployments \cite{gan_open-source_2019}. In this work, we use the social network deployment that provides an online social media application in which users can follow specific users, compose social media posts and view post timelines. This deployment comprises 13 unique MS, supported by  12 in-memory and on-disk data stores. We deploy the MS system on Kubernetes instrumented with Prometheus and Jaeger for metric and trace collection. The social network application provides three different functionalities to users: reading a user's own timeline, reading another user's timeline, or composing a new social media post. To simulate the system under load previous research generally use the provided \textit{wrk2} HTTP workload generator, however, this workload generator only produces static, non-fluctuating, traffic. One of the difficulties of anomaly detection and fault localisation from MSAs is that they are in flux in response to changing workload \cite{luo_characterizing_2021, winchester_temporal_2023}. Therefore, we provisioned the social media application with \textit{Locust} \cite{noauthor_locustio_nodate} instead and generated time-varying workload behaviour based on the experimental design described in \cite{ikram_root_2022}. Specifically, workloads were generated as the sum of sinusoidal waves with randomly generated frequencies in the range $0.01 Hz \leq f \leq 0.06 Hz
$. In all cases, traces were collected in a probabilistic manner at a rate of 0.5\% and metrics were collected at a fixed sampling period of 5s from all MS. To improve the realism of the deployment we used \textit{minikube} \cite{minikube} to deploy the MSA across 5 virtualised worker nodes each with 2 CPU cores and 8000 GB of memory.

To cover a wide array of different fault conditions for evaluation in Section \ref{sec:anomaly_detection_eval} and \ref{sec:rca_eval} we injected three different types of faults commonly employed within the literature: CPU hog, memory leak, and network delay. Memory leak was injected in two different manners, either with or without MS-specific resource usage limits. In the latter case, rather than being throttled the injected MS utilises most of the available memory on its deployed node, causing resource contention with other co-habiting MSs. This fault represents the memory leak noisy neighbour (NN) condition. Each fault was injected into either the compose post, home timeline, user timeline, or user mention MSs. To create each fault dataset the system was first run with time-varying load for 1hr before fault injection, after which the system runs in the degraded state for another hour before observability data collection. Each condition was repeated 5 times, producing 64 datasets in total, in all cases results in Section \ref{sec:anomaly_detection_eval} and \ref{sec:rca_eval} represent the averages across all collected datasets.

\begin{figure*}[ht]
    \centering
    \includegraphics[scale=0.7]{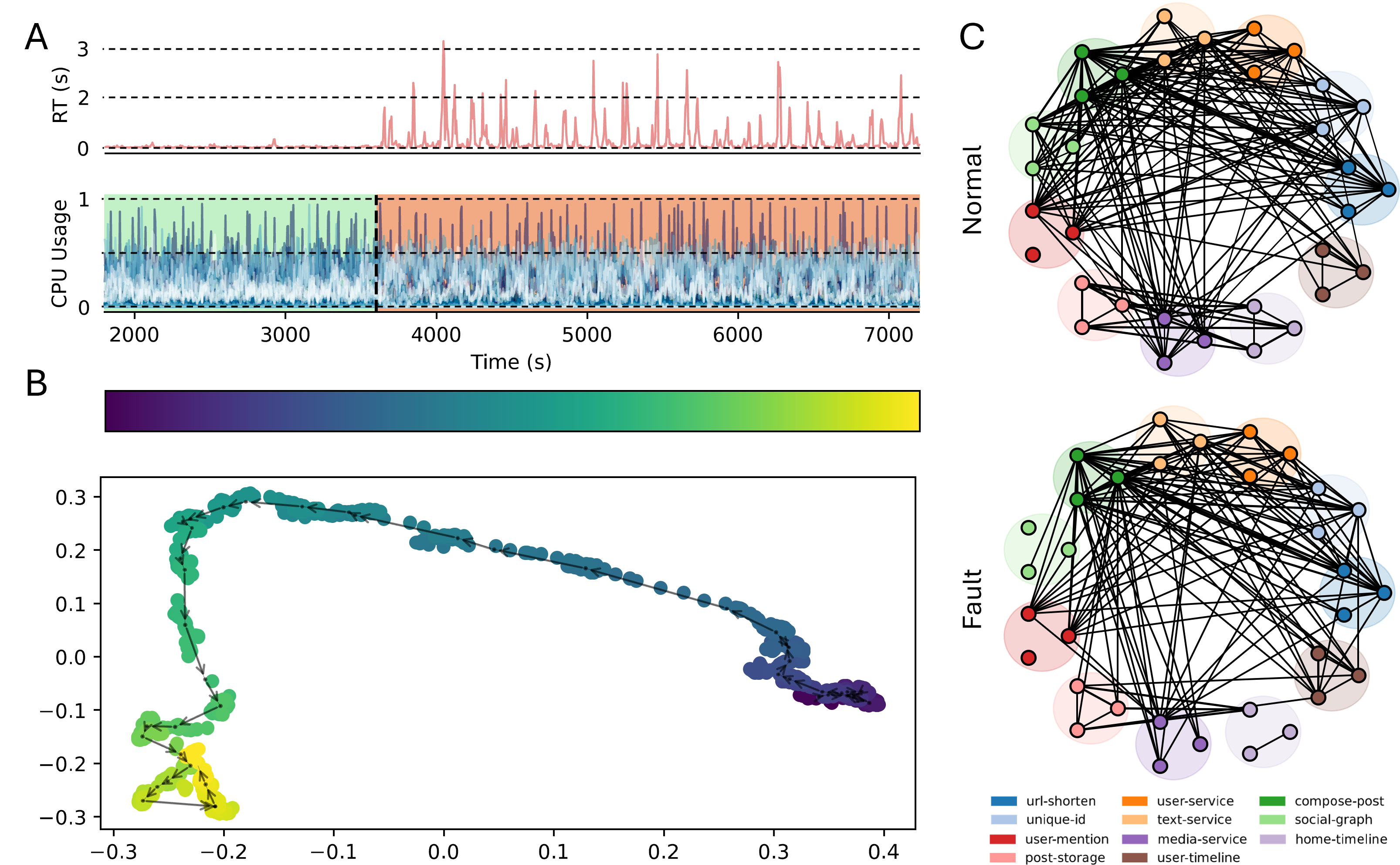} 
    \caption{Panel A: Average response time of HTTP requests (top) and CPU utilsation (bottom) over the course of the experiment with background colour indicating normal (green) and faulty (orange) operation. Panel B: The LMDS 2D embedding of spectral distances between FC graphs, with colour indicating temporal ordering. Panel C: Centroids representing the average FC graph associated with the normal and faulty states of the system.}
    \label{fig:detecting_faults}
\end{figure*}

\subsubsection{Alibaba MS Dataset}
The Alibaba 2022 MS dataset provides runtime metric data during 14 days of operation for a large-scale MS deployment on Alibaba's production cluster supporting a wide array of front-end services \cite{luo_characterizing_2021}. The dataset provides CPU usage information at a MS level with a sampling rate of 60s. However, the dataset does not provide any ground truth as to anomalies that may have occurred during the 13 days of collection, nor what services are being provided. Nonetheless, it provides a good basis in which to benchmark the scalability of detection and localisation approaches.

\subsection{Detection of Time-Varying Changes in MS Dependencies}
\label{sec:time_varying}

In this section, we visualise FC-ADL's ability to capture time-varying changes in dependencies between MSs by inferring and quantifying FC structure rather than focusing on individual metric values.

We generated mixed user traffic for 2.5hrs. For the first 3600s, there was a strict preference for only read home timeline behaviour over user timeline and compose post behaviour. After this initial period, the preference for read home timeline behaviour gradually decreased while the preference for compose post behaviour was increased for a 3600s period. This simulates changing user behaviour at a cohort level. At $T=7200s$, we stopped varying behaviour preferences and instead only increased the number of requests, simulating a spike in traffic, creating a similar pattern of increasing CPU usage but without any underlying change in relationships.

We applied FC-ADL and used landmark multi-dimensional scaling (LMDS) \cite{de_sparse_2004} to embed the resultant distance matrix into 2-dimensional space for visualisation, Figure \ref{fig:time-varying}. The projection demonstrated that the FC graphs showed an initial jump, representing the change from only read home timeline behaviour to mixed behaviour, before following a gradual trajectory in space, aligning temporally with the gradual changes in user behaviour. When user behaviour changes stopped but amplitude increased, the velocity of changes within the system slowed dramatically, with only a slight drift halting almost altogether by the end of the experiment. To some extent, this drift could reflect an artifact of using a sliding windowed approach. Nevertheless, these results demonstrate that our approach can accurately uncover time-varying dependency changes between MS, and highlight when visually observable changes in metric data do not actually imply changes in correlation structure. Approaches that only inspect statistical changes in metrics in isolation would not be able to differentiate between true changing behaviour and demand spikes as they only consider the changes in the absolute values of single metrics. Importantly, FC-ADL does not assume \textit{a priori} information about whether a change indicates a fault, in this case the change in user behaviour may or may not represent an anomaly depending on the cause of the change. Rather FC-ADL detects time-varying changes in dependency structure and informs operators as to what has caused these changes, providing actionable insights if required.

\begin{figure*}[ht]
    \centering
    \includegraphics[width=0.75\textwidth]{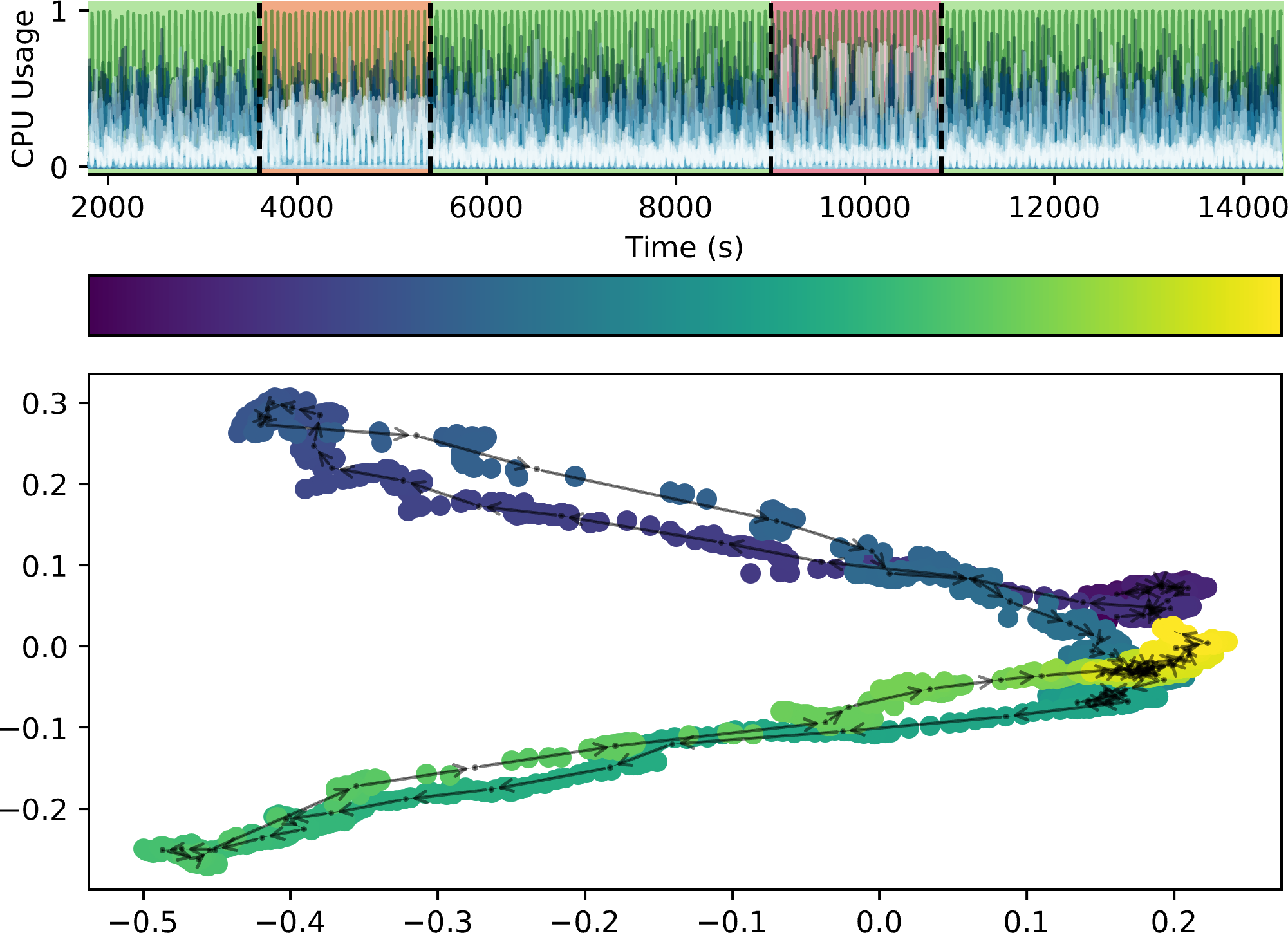} 
    \caption{Panel A: CPU usage for each individual MS instance over the course of the experiment, with background colour indicating normal (green) and faulty (orange) operation. Panel B: The LMDS 2D embedding of spectral distances between FC graphs, with colour indicating temporal ordering.}
    \label{fig:diverse_faults}
\end{figure*}

\subsection{Detecting Faulty Behaviour}
\label{sec:detecting_faults}
In this section, we visually explore whether faulty behaviour in the MS system can be captured through changes in the correlation structure of MS CPU usage and whether these changes could provide informative feedback to network operators. For the purpose of this experiment, we simulate a bottle-neck fault by injecting faults into two of three instances of the home timeline service that caused them to lose functionality.

First, we generated mixed user traffic for 2 hours. For the first hour all MSs operated normally. However, after the first hour a fault was injected within two instances of the home timeline service, resulting in increased system latency (Figure \ref{fig:detecting_faults}A). Using the CPU usage data from this two hour period, we applied FC-ADL. When embedding the resulting FC graphs into a 2-dimensional space using LMDS, we found two distinct groupings corresponding to normal and faulty periods, with a transition period after fault onset (Figure \ref{fig:detecting_faults}B). This visualisation clearly indicates a change at $T=3600s$, which persists until the end of the recording.

To investigate as to whether the structural changes detected by DeltaCon provide insights into the root cause of the injected fault, we applied HDBSCAN to cluster the inferred FC graphs as outlined in Section \ref{sec:detection}. We found that the resultant clusters grouped the inferred FC graphs based on the corresponding normal and faulty period. When inspecting the centroid FC dependency graph for each cluster, depicted in Figure \ref{fig:detecting_faults}C, we identified clear structural markers for the faulty state. Most notably, we found that the FC dependencies between the faulty instances of the home timeline services and other MSs were notably absent, reinforcing our conjecture that the difference between the normal and abnormal FC dependency structure can be utilised for RCA information.

\begin{table*}
\caption{Precision, Recall, F1 score, and time taken for each approach when applied to the four different fault conditions (best scores and fastest times in bold).}
\label{tab:per_fault_bold}
\setlength{\tabcolsep}{8pt}
\begin{tabular}{llcccc}
\toprule
Fault & Method &  F1 & Precision & Recall & Time (s) \\
\midrule
\multirow[t]{5}{*}{CPU Hog} & N-Sigma & 0.77 & 0.65 & 0.98 & 1.83 \\
 & BIRCH & 0.48 & 0.38 & 0.67 & \textbf{0.01} \\
 & SPOT & 0.72 & 0.56 & \textbf{1.00} & 1.98 \\
 & BARO & 0.72 & 0.77 & 0.73 & 129.77 \\
\cmidrule(lr){2-6}
 & FC-ADL (ours) & \textbf{0.90} & \textbf{0.95} & 0.89 & 0.88 \\
\cline{1-6}
\multirow[t]{5}{*}{Memory Leak} & N-Sigma & 0.80 & 0.67 & 1.00 & 1.83 \\
 & BIRCH & 0.64 & 0.50 & 0.88 & \textbf{0.01} \\
 & SPOT & 0.73 & 0.57 & \textbf{1.00} & 2.27 \\
 & BARO & \textbf{0.84} & \textbf{0.91} & 0.84 & 133.43 \\
\cmidrule(lr){2-6}
 & FC-ADL (ours) & 0.81 & 0.90 & 0.76 & 0.84 \\
\cline{1-6}
\multirow[t]{5}{*}{Memory Leak (NN)} & N-Sigma & 0.74 & 0.59 & \textbf{1.00} & 1.84 \\
 & BIRCH & 0.00 & 0.00 & 0.00 & \textbf{0.01} \\
 & SPOT & 0.72 & 0.56 & \textbf{1.00} & 1.56 \\
 & BARO & \textbf{0.94} & \textbf{1.00} & 0.89 & 128.24 \\
\cmidrule(lr){2-6}
 & FC-ADL (ours) & 0.91 & 0.97 & 0.87 & 0.82 \\
\cline{1-6}
\multirow[t]{5}{*}{Network Delay} & N-Sigma & 0.75 & 0.60 & \textbf{1.00} & 1.82 \\
 & BIRCH & 0.45 & 0.35 & 0.62 & \textbf{0.00} \\
 & SPOT & 0.72 & 0.56 & \textbf{1.00} & 1.90 \\
 & BARO & 0.14 & 0.19 & 0.11 & 126.28 \\
\cmidrule(lr){2-6}
 & FC-ADL (ours) & \textbf{0.82} & \textbf{0.85} & 0.82 & 0.86 \\
\cline{1-6}
\bottomrule
\end{tabular}
\label{table:detection}
\end{table*}

\subsection{Detecting Multiple Faulty Behaviours}
\label{sec:diverse_faults}

The previous section showed the ability of our approach to detect transitions between healthy and faulty states for a single fault type. However, real world MS systems are susceptible to a wide array of faults. In this section, we examine FC-ADL's ability to detect transitions between different fault states. We generated mixed traffic for 4 hours with all MSs operating as normal for the first hour. After this period, a transient bottleneck fault was injected into the home timeline service for 30 minutes. After a further hour of normal operation, a transient bottleneck fault was instead injected into the compose post service. Again, this fault was set to last for just 30 minutes.

When applying our framework and using LMDS to visualise the resultant distance matrix in 2-dimensional space, we observed distinct trajectories for the two faults, Figure \ref{fig:diverse_faults}. During the first fault, the system transitioned from the healthy cluster to a cluster representing the home timeline fault, before returning to the healthy cluster. When the second fault was triggered, the system made a second transition from the healthy cluster to another, distinct, cluster representing the compose post fault, again returning to the healthy cluster upon fault remediation. The separation of these fault clusters suggests our approach can distinguish between different types of faults, enabling cluster-based fault labelling. This could enhance the diagnostic capabilities of the approach by enabling quicker identification and resolution of issues based on historical responses to the same fault.

\subsection{Anomaly Detection Evaluation}
\label{sec:anomaly_detection_eval}

To evaluate the effectiveness of FC-ADL in detecting anomalous behaviour indicative of faults within MSAs we follow the steps outlined in previous work \cite{chen_causeinfer_2019, pham_baro_2024}, treating detection models as binary classifiers. The metric data collected across all fault conditions outlined in Section \ref{sec:experimental_setup} is labelled as normal for data points that occurred prior to fault injection time, and abnormal for all data points after. Similarly to \cite{pham_baro_2024} for all benchmark algorithms and FC-ADL the first detected change-point is taken as the anomaly decision boundary, labelling all prior points as normal and all subsequent points as abnormal. Predictions can then be compared to the ground-truth for each dataset to find True Positives (TP) based on correct abnormal labelling, True Negatives (TN) based on correct normal labelling, False Positives (FP) based on incorrect abnormal labelling, and False Negatives (FN) based on incorrect normal labelling. These values can then be used to calculate the Precision, Recall, and F1 scores for each approach.

\subsubsection{Baselines}

For anomaly detection we select four baselines: N-Sigma \cite{li_causal_2022, lin_microscope_2018}, BIRCH \cite{wu_microdiag_2021, xin_causalrca_2023}, SPOT \cite{lee_eadro_2023, li_causal_2022, meng_localizing_2020}, and Multivariate Bayesian Online Change point Detection (MBOCPD) as implemented by BARO \cite{pham_baro_2024}. These approaches have been selected due to their common use for anomaly detection RCA research as well as state-of-the-art performance \cite{pham_baro_2024}.

\begin{itemize}
    \item \textbf{N-Sigma}: is a simple anomaly detection method utilised across many RCA works for fault detection \cite{lin_microscope_2018, yu_microrank_2021, li_causal_2022, lee_eadro_2023}. In this work we use the 3-Sigma rule implementation \cite{pham_baro_2024}, being the most common version of N-Sigma, which relies on the fact that the vast majority of data points in a normal distribution fall within three standard deviations of the mean. Therefore, historical data is used to estimate the mean and standard deviations of metrics, and then for anomaly detection anything that falls outside of the three standard deviation threshold during runtime is flagged as anomalous.
    \item \textbf{BIRCH}: this method of anomaly detection has been implemented in prior RCA work \cite{wu_microdiag_2021, xin_causalrca_2023} and uses BIRCH clustering applied to historical, normal, metric values \cite{wu_microdiag_2021, xin_causalrca_2023}. In this paper we use the implementation of BIRCH described in \cite{pham_baro_2024} in which BIRCH considers a point to be anomalous when it is assigned a different cluster with consecutive data points.
    \item \textbf{SPOT}: is an approach based off of Extreme Value (EV) Theory \cite{siffer_anomaly_2017}. In this work we use the dSPOT variant of SPOT used in \cite{meng_localizing_2020} and previous benchmarks \cite{pham_baro_2024}, hereafter any references to SPOT refer to the dSPOT variant. This variant of SPOT is designed to handle shifts in the underlying time series by first removing local trends via a sliding-window model and maintaining a dynamic EV threshold based on iterative re-fitting. Values that exceed the EV threshold are considered anomalous.
    \item \textbf{BARO}: carries out anomaly detection via Multivariate Bayesian Change Point Detection, an approach that models the number of consecutive points since the last distribution change to detect change points in multivariate time series metrics. Unlike the above approaches, but similar to our approach, BARO inherently captures the relationships between metrics making it more robust for change point detection from MSAs \cite{pham_baro_2024}.
\end{itemize}

\subsubsection{Results}

Our results in Table \ref{table:detection} demonstrate that FC-ADL outperformed N-Sigma, BIRCH, and SPOT in all cases in terms of F1 score. When compared to BARO, we find that FC-ADL performs similarly for fault detection, having slightly higher F1 scores in two cases (network delay and memory leak) and slightly lower F1 scores in two others (CPU hog and memory leak noisy neighbour). However, BARO's runtime was significantly higher than any other anomaly detection approach, being 152x slower than FC-ADL on average. The high computational overhead of BARO comes from the the inherent scalability limitations of BOCPD discussed in Section \ref{sec:introduction} as the number of variables and number of samples increases. Overall, these results demonstrate that FC-ADL delivers state-of-the-art accuracy in detecting the onset of anomalous MSA behaviour from FC inferred from MS metrics, while remaining highly efficient when compared to BARO.

Whilst N-Sigma and SPOT consistently demonstrated the highest recall results their overall F1 scores were significantly affected by their poor precision due to routinely predicting anomaly onsets early. These results are likely a product of our more realistic experimental setup with varying user load (see Section \ref{sec:experimental_setup}). Without consideration for the relationship between metrics, like with BARO and FC-ADL, these methods are sensitive to changes in metric values caused by normal changes in user demand. Overall, these findings illustrate the importance of developing anomaly detection approaches for MSAs that consider their inherent inter-dependent nature and testing approaches in more realistic and time-varying environments.

One interesting finding from our evaluation is that faults not immediately directly affecting CPU usage, such as memory leaks and network delay, can still be accurately detected using only CPU usage data, especially in the case of BARO and FC-ADL. This indicates that subtle shifts in CPU usage reflect fault-induced performance changes and coupling between microservices. Consequently, this raises questions about the minimal variety of metric sources needed for effective anomaly detection, potentially enabling more efficient observability.

\subsection{Root Cause Localisation Evaluation}
\label{sec:rca_eval}

\begin{table*}
\caption{Average top-1, top-3, and top-5 accuracy alongside corresponding computation time for each approach when applied to the four different fault conditions (best scores and fastest times in bold).}
\label{tab:rca_comparison}
\setlength{\tabcolsep}{8pt}
\begin{tabular}{llcccc}
\toprule
Fault & Method & Avg@1 & Avg@3 & Avg@5 & Time (s)  \\
\midrule
\multirow[t]{7}{*}{CPU Hog} & $\epsilon$-Diagnosis & 0.00 & 0.08 & 0.15 & 0.70 \\
 & PC+PR & 0.00 & 0.00 & 0.00 & 4.93 \\
 & DirectLiNGAM+PR & 0.06 & 0.06 & 0.06 & 1.63 \\
 & CIRCA & 0.71 & 0.83 & \textbf{0.92} & 3.06 \\
 & N-Sigma & 0.00 & 0.06 & 0.15 & \textbf{0.01} \\
 & BARO & 0.06 & 0.44 & 0.77 & 0.03 \\
\cmidrule(lr){2-6}
 & FC-ADL (ours) & \textbf{0.85} & \textbf{0.85} & 0.85 & 0.02 \\
\cline{1-6}
\multirow[t]{7}{*}{Memory Leak} & $\epsilon$-Diagnosis & 0.06 & 0.19 & 0.19 & 0.70 \\
 & PC+PR & 0.00 & 0.00 & 0.00 & 4.79 \\
 & DirectLiNGAM+PR & 0.00 & 0.06 & 0.06 & 1.64 \\
 & CIRCA & \textbf{0.81} & \textbf{0.94} & \textbf{1.00} & 3.10 \\
 & N-Sigma & 0.12 & 0.12 & 0.19 & \textbf{0.00} \\
 & BARO & 0.00 & 0.44 & 0.88 & 0.02 \\
\cmidrule(lr){2-6}
 & FC-ADL (ours) & 0.68 & 0.81 & 0.90 & 0.02 \\
\cline{1-6}
\multirow[t]{7}{*}{Memory Leak (NN)} & $\epsilon$-Diagnosis & 0.00 & 0.06 & 0.25 & 0.70 \\
 & PC+PR & 0.00 & 0.00 & 0.00 & 1.27 \\
 & DirectLiNGAM+PR & 0.00 & 0.00 & 0.00 & 1.65 \\
 & CIRCA & 0.00 & 0.19 & 0.38 & 1.99 \\
 & N-Sigma & 0.00 & 0.12 & 0.38 & \textbf{0.01} \\
 & BARO & 0.00 & 0.00 & 0.25 & 0.02 \\
\cmidrule(lr){2-6}
 & FC-ADL (ours) & \textbf{0.10} & \textbf{0.34} & \textbf{0.55} & 0.03 \\
\cline{1-6}
\multirow[t]{7}{*}{Network Delay} & $\epsilon$-Diagnosis & 0.19 & 0.19 & 0.25 & 0.70 \\
 & PC+PR & 0.00 & 0.12 & 0.19 & 5.62 \\
 & DirectLiNGAM+PR & 0.00 & 0.12 & 0.12 & 1.64 \\
 & CIRCA & 0.00 & 0.31 & 0.38 & 2.89 \\
 & N-Sigma & 0.00 & 0.00 & 0.06 & \textbf{0.01} \\
 & BARO & 0.00 & 0.00 & 0.00 & 0.01 \\
\cmidrule(lr){2-6}
 & FC-ADL (ours) & \textbf{0.28} & \textbf{0.50} & \textbf{0.51} & 0.02 \\
\cline{1-6}
\bottomrule
\end{tabular}
\label{table:rca}
\end{table*}

To evaluate the capability of FC-ADL to accurately identify the root cause of faults we compare its performance to several baselines in extracting the top-K most likely root cause from the datasets discussed in Section \ref{sec:experimental_setup}. All baselines and FC-ADL were applied to the full dataset for anomaly detection and subsequent RCA. For baselines that do not include anomaly detection (CIRCA, $\epsilon$-Diagnosis), the ground-truth fault onset time is given instead as input. For each type of fault (CPU hog, memory leak, memory leak noisy neighbour, and network delay) we report the average top-K accuracy, indicating how often each method listed the true root cause MS that received the injected fault in the top 1, 3, or 5 candidate faulty MS(s). In each case, we also report the average time taken for exclusively RCA.

\subsubsection{Baselines.}
For RCA we select six baseline approaches: $\epsilon$-Diagnosis \cite{shan_diagnosis_2019}, N-Sigma \cite{li_causal_2022, lin_microscope_2018}, CIRCA \cite{li_causal_2022}, PC \cite{spirtes_causation_1993}, LiNGAM \cite{shimizu_linear_2006, shimizu_directlingam_2011}, and BARO \cite{pham_baro_2024}. As in Section \ref{sec:anomaly_detection_eval}, these approaches were selected due to their open-source implementations, state-of-the-art performance, and use in previous work as benchmarks.

\begin{itemize}
    \item \textbf{$\epsilon$-Diagnosis}: an unsupervised RCA method that utilises non-parametric two-sample hypothesis testing using $\epsilon$-statistics \cite{shan_diagnosis_2019} to quantify how much the distribution of metrics between normal and abnormal data has changed into a p-value. Metrics whose p-value falls below a confidence threshold are flagged as candidate root causes and are ranked by total p-value.

    \item \textbf{N-Sigma}: carries out RCA by comparing the mean and standard deviation of each metric over the normal data to the detected abnormal period \cite{li_causal_2022, lin_microscope_2018}. Each metric is then scored based on the maximum z-score, quantifying how far each value is from its pre-fault average. Each metric is ranked by descending score to give the top candidate root causes.

    \item \textbf{CIRCA}: creates a causal graph by using tracing data and domain knowledge \cite{li_causal_2022}. As in prior work \cite{pham_baro_2024} we supply a causal graph constructed by the PC algorithm from metrics based on the normal functioning of the system. A regression model is then trained on normal data to estimate its conditional distribution. The model can then be applied to the anomalous data to quantify how much each metric's values deviate from the normal period. Candidate root causes are then ranked based on their total deviation.
    
    \item \textbf{PC} \cite{spirtes_causation_1993}: a popular approach for constructing causal graphs between metrics is the PC algorithm \cite{ma_automap_2020, xin_causalrca_2023, ma_ms-rank_2019, ma_servicerank_2022, lin_facgraph_2018} and thus is a common baseline in prior research \cite{xin_causalrca_2023, pham_baro_2024}. Following the design of MicroDiag \cite{wu_microdiag_2021} we provide edge-weights based on the correlation between metrics and inverse directed edges before applying PageRank to list the top-K root causes.

    \item \textbf{LiNGAM} \cite{shimizu_linear_2006, shimizu_directlingam_2011}: another popular approach for constructing causal graphs between metrics is the LiNGAM algorithm \cite{wu_microdiag_2021}. Similarly to above we follow MicroDiag \cite{wu_microdiag_2021} by providing edge-weights based on the correlation between metrics and inverse directed edges before applying PageRank to list the top-K root causes.

    \item \textbf{BARO}: similarly to $\epsilon$-Diagnosis and N-Sigma, BARO carries out fault localisation via non-parametric hypothesis testing using their \textit{Robust Scorer}, comparing the statistical distributions of the detected normal and abnormal data. Unlike previous approaches, \textit{Robust Scorer} uses the median and interquartile range which are more robust to outliers making the approach more resilient to imprecise anomaly time onset estimates.
\end{itemize}

\subsubsection{Results}

Our results in Table \ref{table:rca} demonstrate that across all faults FC-ADL outperformed, or performed on-par with, benchmark approaches in accurately identifying candidate root causes. Whilst most approaches struggled with identifying the true root cause in the top-1 of candidates, both FC-ADL and CIRCA managed to achieve impressive accuracy for CPU hog and memory leak faults. Most approaches saw improved top-3 and top-5 accuracy across all fault types, notably, FC-ADL performed above all baselines in accurately identifying the root cause of memory leak (NN) and network delay faults. Whilst CIRCA demonstrated impressive performance across all fault types the approach does not provide end-to-end fault detection and localisation, and thus ground truth injection time was supplied. In practice, it's almost impossible to pinpoint the exact moment a fault occurs so as to cleanly partition the metrics into pre-fault and post-fault sets, and the resulting data contamination is likely to impair the accuracy of the logistic regression model. The performance of BARO in identifying the true root cause in the top-1 and top-3 candidates was unexpectedly poor given its excellent performance in Section \ref{sec:anomaly_detection_eval}, however, across CPU Hog and memory leak faults BARO saw notable improvements in top-5 accuracy to near comparable performance of CIRCA and FC-ADL. The overall poor performance of many baseline approaches for RCA on the generated datasets with time-varying demand and thus non-stationary metrics may indicate the necessity to test such approaches on more realistic environments to better understand their potential real-world performance.

Our results showed that most approaches saw significant reductions in performance in accurately locating memory leak (NN) and network delay faults despite these same reductions not being substantially noticeable in the evaluation of anomaly detection in Section \ref{sec:anomaly_detection_eval}. As this same decrease was not observed in the normal memory leak scenario, it suggests that the resource contention, and thus indirect fault propagation with co-located MSs, in the memory leak NN scenario may mask performance signatures that RCA approaches rely on. As indirect fault propagation through cohabitation is expected within real-world MSAs \cite{ma_automap_2020}, but is under-represented in RCA test bed research, future work should consider such fault propagation when benchmarking their approaches to better understand their performance under diverse fault conditions. The decrease in performance for network delay faults may indicate that whilst CPU usage in isolation can help accurately identify the onset of an anomaly, it may not be sufficient in all cases to accurately identify the root cause.

Whilst the computational overhead of all approaches was low in our experiments, we note that even at such a small scale the computational efficiency of FC-ADL is noticeable. Overall, FC-ADL demonstrated a 202x and 75x speed up compared to PC and DirectLiNGAM respectively.

\begin{figure}[h]
    \centering
    \includegraphics[scale=0.85]{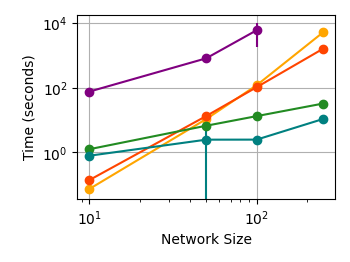}
    \caption{Scalability of BARO (purple), the PC algorithm (yellow), DirectLiNGAM (orange), N-Sigma (green), and FC-ADL (teal) to the Alibaba deployment.}
    \label{fig:rca_scalability}
\end{figure}

\begin{figure}[h]
    \centering
    \includegraphics[scale=0.8]{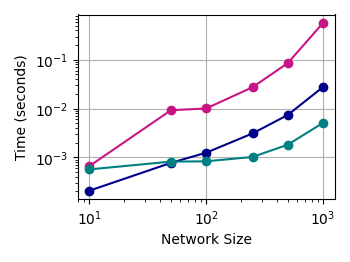} 
    \caption{Scalability of each component of FC-ADL, EWMA inference (blue), DeltaCon (pink), and HDBSCAN (teal) for a single window.}
    \label{fig:scalability}
\end{figure}

\subsection{Scalability Analysis}
\label{sec:scalability}

To illustrate FC-ADL's scalability, we used a real-world dataset released by Alibaba \cite{luo_power_2022} to test the scalability of FC-ADL and other baseline RCA approaches. As there is no ground-truth information in the dataset as to the onset of faults, we only include RCA approaches that do not require specific fault onset timings. We applied each approach to increasingly large numbers of randomly sampled MSs from the Alibaba dataset using 1080 samples of CPU usage data from each MS, corresponding to 12 hours of recordings. We only demonstrate up to 250 MSs, and only 100 MSs for BARO, due to the extreme computational overheads of approaches beyond these points. Our results in Figure \ref{fig:rca_scalability}, demonstrate that for just 100 nodes BARO's end-to-end approach exceeds processing times of 1hr. Whilst DirectLiNGAM and PC perform better, they still near 1hr processing times when applied to just 250 MSs. On the other hand, FC-ADL took 3s for 100 MSs and just 11s for 250 MSs to provide actionable RCA information. It is worth noting that N-Sigma achieved similar performance to that of FC-ADL, however, we observe that N-Sigma performs poorly in both anomaly detection and localisation accuracy (Sections \ref{sec:anomaly_detection_eval} and \ref{sec:rca_eval}. Therefore, it is unclear as to how valuable the insights produced by N-Sigma would be despite its efficiency. These results highlight the inability of prior causal inference and multivariate based approaches to scale to large-scale deployments, and demonstrates the comparative computational efficiency of FC-ADL.

One of the benefits of FC-ADL is that, unlike other approaches evaluated in this paper \cite{pham_baro_2024, wu_microdiag_2021, li_causal_2022, li_causal_2022, lin_microscope_2018, shan_diagnosis_2019}, it can be applied as data arrives to construct new FC-based dependencies in an online manner. Our results in Figure \ref{fig:scalability} demonstrate that when FC-ADL is applied to construct a single window, calculate DeltaCon distance, and use HDBSCAN to predict the cluster, the approach remains sub-second even up to 1000 MSs. This further reinforces the scalability of FC-ADL when applied to large-scale MSAs, highlighting its suitability for real-time, incremental fault localisation where continuous monitoring and rapid insights are essential.

Utilising only metric data comes with additional benefits of reduced data storage and collection costs. The uncompressed size on disk of both CPU and memory usage data for all MS instances, around 18x greater than the number of unique MS, over a day is 52.3GB after post-processing. On the other hand, the size on disk of distributed tracing data, only for unique MSs and sampled at just 0.5\%, is 713.5GB after post-processing. This signifies a 92.7\% reduction in data size whilst retaining a much higher spatial granularity of data at an instance-level.

\section{Conclusion}
In this paper we have introduced FC-ADL, a framework for monitoring time-varying dependencies within MS systems by applying FC to MS usage data. We demonstrated that by tracking changes in these dependencies, FC-ADL can accurately detect the occurrence of anomalies. Furthermore, we demonstrated that by examining the dependency structure before and after a fault has occurred, the most likely candidate cause for the anomaly can be found. We demonstrated that overall, FC-ADL can achieve state-of-the-art performance in both anomaly detection and RCA on a wide variety of fault conditions and MSs, and does so with low computational overhead.

A design choice prevalent throughout FC-ADL is a focus on scalability, imparted due to the usage of efficient correlation calculations to infer the time-varying FC between MSs. This efficiency enables FC-ADL to be deployed to large-scale MSAs to provide timely fault detection and localisation information, which is difficult or impossible with prior work. One important contribution of FC-ADL is demonstrating that causation is not necessarily required for accurate fault localisation. Indeed, we suggest that correlation-based approaches may be preferred in cases where timely fault localisation is paramount, especially in cases where the number of variables is very large or when fault data is limited due to decreased complexity of modelling.

Whist FC-ADL is indeed computationally efficient, the use of Pearson's correlations can result in somewhat noisy FC inference. This can mean that more data is required per window to infer meaningful correlations, which can then mean that the framework is slower to adapt to changes in dependency structure, even when mitigated by a EWMA. Future iterations of FC-based anomaly detection and RCA within MSAs could explore alternative, more computationally demanding, statistical measures that can provide more accurate and robust measures of functional coupling \cite{friston_functional_2011}. Another avenue for future work is to explore the use of FC-based prior knowledge to reduce the computational overhead of causal inference approaches, which has been demonstrated to achieve significant speed-ups when applied to non-MSA architectures \cite{winchester_accelerating_2022}. These avenues could provide options fora  trade-off between fast and efficient RCA and potentially more accurate RCA, providing a hierarchical triage framework.


\end{document}